\documentclass{ws-ijmpcs}

\usepackage{graphicx}
\usepackage{amssymb}
\usepackage{amsmath}    
\usepackage{txfonts}
\usepackage{xspace}

\newcommand{\Chandra}{\textit{Chandra}\xspace}

\newcommand{\FermiLAT}{\textit{Fermi}-LAT\xspace}

\newcommand{\He}{H.E.S.S.\xspace}
\newcommand{\Ve}{VERITAS\xspace}

\newcommand{\Ma}{MAGIC\xspace}

\newcommand{\M}{M\,87\xspace}

\begin{document}

\markboth{M. Raue et al.}
{The 2010 \M VHE flare \& the MWL picture}

%
\catchline{}{}{}{}{}
%

\title{THE 2010 M 87 VHE FLARE AND ITS ORIGIN: THE MULTI-WAVELENGTH PICTURE}

\author{
M. RAUE\footnote{Institut f\"ur Experimentalphysik, Universit\"at Hamburg, Hamburg, Germany, \texttt{martin.raue@desy.de}},
L.~STAWARZ\footnote{Obserwatorium Astronomiczne, Uniwersytet Jagiello{\'n}ski, ul. Orla 171, 30-244 Krak{\'o}w, Poland},
D.~MAZIN\footnote{IFAE, Edifici Cn., Campus UAB, E-08193 Bellaterra, Spain},
P.~COLIN\footnote{Max-Planck-Institut f\"ur Physik, D-80805 M\"unchen, Germany},
C.~M.~HUI\footnote{Department of Physics and Astronomy, University of Utah, Salt Lake City, UT 84112, USA},
M.~BEILICKE\footnote{Department of Physics, Washington University, St. Louis, MO 63130, USA},\\
W.~MCCONVILLE\footnote{NASA Goddard Space Flight Center, Greenbelt, MD 20771, USA, and Department of Physics and Department of Astronomy, University of Maryland, College Park, MD 20742, USA},
\;M.~GIROLETTI\footnote{INAF Istituto di Radioastronomia, 40129 Bologna, Italy},
\;D.~E.~HARRIS\footnote{Smithsonian Astrophysical Observatory, 60 Garden St., Cambridge, MA 02138, USA},
\setcounter{footnote}{0}
I.~A.~STEELE\footnote{Astrophysics Research Institute, Liverpool John Moores University, UK},
R.C.~WALKER\footnote{National Radio Astronomy Observatory (NRAO), Socorro, NM 87801, USA},\\
FOR THE H.E.S.S., MAGIC, VERITAS, AND FERMI-LAT COLLABORATIONS AND\\ THE \M MWL MONITORING TEAMS}

\maketitle

\begin{history}
\received{Day Month Year}
\revised{Day Month Year}
\end{history}

\begin{abstract}
The giant radio galaxy \M, with its proximity (16\,Mpc) and its very massive black hole ($(3-6) \times10^9 M_\odot$), provides a unique laboratory to investigate very high energy (E$>$100\,GeV; VHE) gamma-ray emission from active galactic nuclei and, thereby, probe particle acceleration to relativistic energies near supermassive black holes (SMBH) and in relativistic jets. \M has been established as a VHE $\gamma$-ray emitter since 2005. The VHE $\gamma$-ray emission displays strong variability on timescales as short as a day. In 2008, a rise in the 43\,GHz Very Long Baseline Array (VLBA) radio emission of the innermost region (core; extension of $<100 R_{\rm s}$; Schwarzschild radii) was found to coincide with a flaring activity at VHE. This had been interpreted as a strong indication that the VHE emission is produced in the direct vicinity of the SMBH. In 2010 a flare at VHE was again detected triggering further multi-wavelength (MWL) observations with the VLBA, Chandra, and other instruments. At the same time, \M was also observed with the \FermiLAT telescope at MeV/GeV energies, the European VLBI Network (EVN), and the Liverpool Telescope (LT). Here, preliminary results from the 2010 campaign will be reported.

\end{abstract}

\ccode{PACS numbers: 98.54.Gr, 95.85.Pw, 95.85.Nv, 95.85.Kr, 95.85.Bh}

\section{Introduction}

\begin{figure}[tb]
\centering
\includegraphics[width=0.5\textwidth]{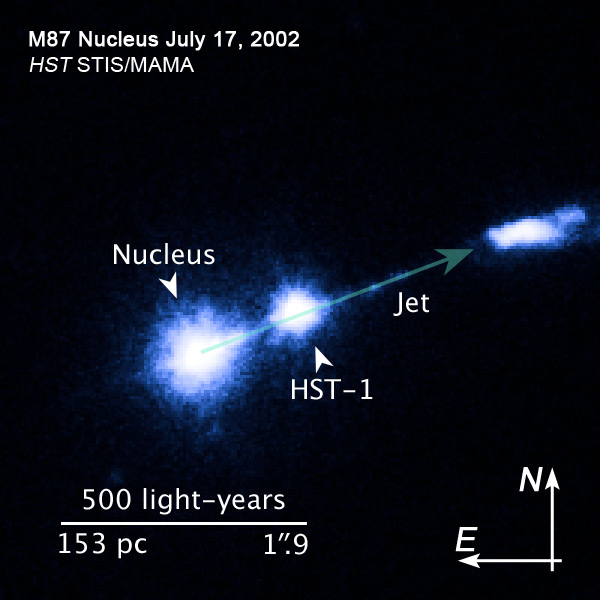}
\caption{Hubble Space Telescope (HST) image of M\,87. \textit{(Illustration Credit: NASA, ESA, and Z. Levay, STScI; Credit: NASA, ESA, and J. P. Madrid, McMaster University)} \label{Fig:M87HST}}
\end{figure}

The giant radio galaxy \M provides a unique environment to study relativistic plasma outflows and the surrounding of supermassive black holes (SMBH). Its prominent jet \cite{curtis:1918a:m87jet} is resolved from radio to X-rays, displaying complex structures (knots, diffuse emission, \cite{perlman:1999a,perlman:2001a}), strong variability \cite{harris:2003a,harris:2006a}, and superluminal motion \cite{biretta:1999a,cheung:2007a} (Fig.~\ref{Fig:M87HST}\cite{madrid:2009a}). With its proximity ($16.7 \pm 0.2$\,Mpc; \cite{mei:2007a}) and its very massive black hole of $M_{\rm BH} \simeq (3-6) \times 10^{9}$\,M$_\odot$ \cite{macchetto:1997a,gebhardt:2009a}
high-resolution radio observations enable one to directly probe structures with sizes down to $<200$ Schwarzschild radii.

\M has been established as a very high energy (VHE; $E>100$\,GeV) emitter since 2005 \cite{aharonian:2003b,aharonian:2006:hess:m87:science}. It is one of only four extragalactic VHE sources belonging to the class of radio galaxies for which only weak or moderate beaming of the emission is expected. \M shows strong variability at VHE with timescales of the order of days \cite{aharonian:2006:hess:m87:science,albert:2008:magic:m87,acciari:2009b:m87joinedcampaign:science}. This indicates a compact emission region $< 5 \times 10^{15} \delta$\,cm ($\delta$: bulk Doppler factor of the emitting plasma), corresponding to only a few tens of Schwarzschild radii. \M has recently been detected at GeV energies by \FermiLAT \cite{abdo:2009:fermi:lat:m87}.

The angular resolution of ground-based VHE instruments\footnote{Typically, $\sim$0.1\,degree per event, corresponding to $\sim30$\,kpc projected size.} does not allow for a direct determination of the origin of the VHE emission in the inner kpc structures. To further investigate the location of the VHE emission site and the associated production mechanisms, variability studies and the search for correlations with other wavelengths need to be utilized \cite{acciari:2009b:m87joinedcampaign:science}.

Up to now, three episodes of enhanced VHE activity have been detected from \M. The first one, in 2005 \cite{aharonian:2006:hess:m87:science}, occurred during an extreme radio/optical/X-ray outburst of the jet feature HST-1 \cite{harris:2003a,harris:2006a}, which has been discussed as a possible site for the VHE emission \cite{stawarz:2006a,cheung:2007a,harris:2009a}. During the second flaring episode, in 2008, HST-1 was in a low flux state, but radio measurements showed a flux increase of the core region within a few hundred Schwarzschild radii of the SMBH, suggesting the direct vicinity of the SMBH as the origin of the VHE emission \cite{acciari:2009b:m87joinedcampaign:science}. This conclusion was further supported by the detection of an enhanced X-ray flux from the core region by \Chandra.

The third episode of increased VHE activity occurred in 2010 during a joint VHE monitoring campaign by \Ma and \Ve. The detection of the high state \cite{mariotti:2010:magic:m87:atel,ong:2010a:m87:veritas:magic:flare:atel} triggered further multi-wavelength (MWL) observations by the VLBA, \Chandra, and other instruments. Preliminary results from the campaign are presented in this paper, while the final campaign results will be reported in an upcoming publication.

\section{The 2010 VHE flare}

\begin{figure}[tb]
\centering
\includegraphics[width=\textwidth]{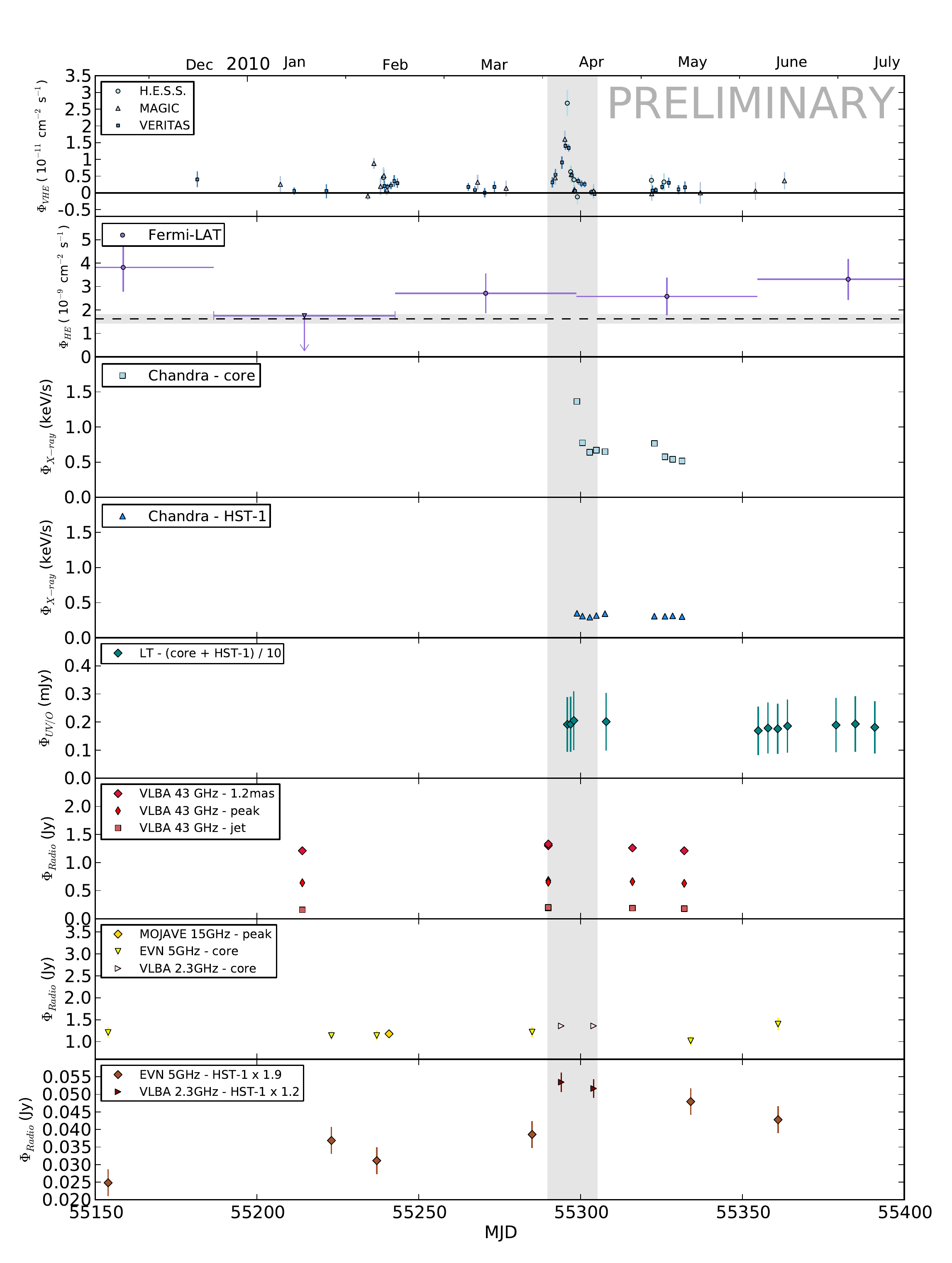}
\caption{Preliminary MWL light curve from the 2010 monitoring campaign on \M. The integral fluxes in the VHE range are shown above an energy of 350\,GeV. The radio flux of HST-1 at different frequencies has been normalized to the 5\,GHz flux assuming a spectrum $S_\nu \sim \nu^{-\alpha}$ with $\alpha = 0.6$. More details on the data sets, analysis, and the characteristics of the different instruments can be found in an upcoming publication.
\label{Fig:MWLLightcurve}}
\end{figure}

The preliminary combined VHE light curve for the 2010 monitoring campaign is shown in Fig.~\ref{Fig:MWLLightcurve} upper panel. During the campaign, two episodes of enhanced VHE $\gamma$-ray emission have been detected \cite{mariotti:2010:magic:m87:atel,ong:2010a:m87:veritas:magic:flare:atel}: The first episode took place in Feb. 2010, where a single night of increased activity was detected by \Ma. VHE follow-up observations did not reveal further activity. The second episode took place in Apr. 2010 and showed a pronounced VHE flare detected by several instruments, triggering further MWL observations.

The VHE activity of this second flaring episode is concentrated in a single observation period between MJD 55290 and MJD 55305 ($\sim$15\,days). This time period is exceptionally well covered by observations with 21 pointings by \He, \Ma, and \Ve, resulting in an observation almost every night. The detected flare displays a smooth rise and decay in flux with a peak around MJD 55296 (April 9-10, 2010; see Fig.~\ref{Fig:MWLLightcurve}). In general, during nights with quasi simultaneous observations by different instruments, the measured fluxes are found to be in excellent agreement.

Compared to previous VHE flares detected in 2005 and 2008, the 2010 flare shows similar timescales and peak flux levels, but the overall variability pattern is somewhat different (more extended periods of flaring activity with several flux maxima), though the statistics and the sampling of the previous VHE flares limit a definitive conclusion. 

\section{2010 multi-wavelength observations}

The discovery of a VHE high state in April 2010 triggered further multi wavelength (MWL) observations, which are displayed in Fig.~\ref{Fig:MWLLightcurve}.

\Chandra started observing two days after the peak VHE flux had been detected, performing five pointings  spaced in intervals between 1.5 and 3 days (5ks each). A second set of four observations was taken starting about two weeks later. HST-1 was found in a low flux state, while the core showed an increase in X-ray flux in the first observation that followed the VHE flare. Further details on the \Chandra observations can be found in Harris et al. in these proceedings.

A total of five VLBA 43\,GHz observations were taken in 2010, three monitoring observations and two additional observations following the detection of the VHE high state. No enhanced radio flux from the core region was detected, contrary to what was observed during the 2008 VHE outburst \cite{acciari:2009b:m87joinedcampaign:science}.

During the 2010 campaign, further MWL data were taken by the Liverpool Telescope in the optical, the EVN, and the VLBA, and \M was continiously monitored at MeV/GeV energies with the \FermiLAT. No significant variability is found in the 2\,year  \FermiLAT data set spanning from August  2008 to August 2010. The details of these observations will be reported in an upcoming paper.

\section*{Acknowledgments}

{\small
M. Raue acknowledges generous support from the LEXI program of the state of Hamburg, Germany.
--- The \He Collaboration acknowledges support of the Namibian authorities and of the University of Namibia
in facilitating the construction and operation of H.E.S.S., as is the support by the German Ministry for Education and
Research (BMBF), the Max Planck Society, the French Ministry for Research,
the CNRS-IN2P3 and the Astroparticle Interdisciplinary Programme of the
CNRS, the U.K. Science and Technology Facilities Council (STFC),
the IPNP of the Charles University, the Polish Ministry of Science and 
Higher Education, the South African Department of
Science and Technology and National Research Foundation, and by the
University of Namibia. We appreciate the excellent work of the technical
support staff in Berlin, Durham, Hamburg, Heidelberg, Palaiseau, Paris,
Saclay, and in Namibia in the construction and operation of the
equipment.
--- The MAGIC Collaboration thanks the Instituto de Astrof'sica de Canarias for the excellent working conditions at the Observatorio del Roque de los Muchachos in La Palma. The support of the German BMBF and MPG, the Italian INFN and Spanish MICINN is gratefully acknowledged.
--- The VERITAS Collaboration acknowledges support from the US Department of Energy, the US National Science Foundation, and the Smithsonian Institution, from NSERC in Canada, from Science Foundation Ireland (SFI 10/RFP/AST2748), and from STFC in the UK. We acknowledge the excellent work of the technical support staff at the FLWO and at the collaborating institutions in the construction and operation of the instrument.
--- The $Fermi$ LAT Collaboration acknowledges support from a number of agencies and institutes for both development and the operation of the LAT as well as scientific data analysis. These include NASA and DOE in the United States, CEA/Irfu and IN2P3/CNRS in France, ASI and INFN in Italy, MEXT, KEK, and JAXA in Japan, and the K.~A.~Wallenberg Foundation, the Swedish Research Council and the National Space Board in Sweden. Additional support from INAF in Italy and CNES in France for science analysis during the operations phase is also gratefully acknowledged.
--- The Very Long Baseline Array is operated by the
National Radio Astronomy Observatory, a facility of the NSF, operated
under cooperative agreement by Associated Universities, Inc.
--- The European VLBI Network is a joint facility of European, Chinese, South African and other radio astronomy institutes funded by their national research councils. This effort is supported by the European Community Framework Programme 7, Advanced Radio Astronomy in Europe, grant agreement No. 227290.
--- This research has made use of data from the MOJAVE database that is maintained by the MOJAVE team \cite{lister:2009a}.
--- This research has made use of NASA's Astrophysics Data System.
}


%
%
\def\Journal#1#2#3#4{{#4}, {#1}, {#2}, #3}
\def\NAT{Nature}
\def\AAA{A\&A}
\def\ApJ{ApJ}
\def\AJ{Astronom. Journal}
\def\Aph{Astropart. Phys.}
\def\ApJS{ApJSS}
\def\MNRAS{MNRAS}
\def\NIM{Nucl. Instrum. Methods}
\def\NIMA{Nucl. Instrum. Methods A}
\def\aj{AJ}%
\def\actaa{Acta Astron.}%
\def\araa{ARA\&A}%
\def\apj{ApJ}%
\def\apjl{ApJ}%
\def\apjs{ApJS}%
\def\ao{Appl.~Opt.}%
\def\apss{Ap\&SS}%
\def\aap{A\&A}%
\def\aapr{A\&A~Rev.}%
\def\aaps{A\&AS}%
\def\azh{AZh}%
\def\baas{BAAS}%
\def\bac{Bull. astr. Inst. Czechosl.}%
\def\caa{Chinese Astron. Astrophys.}%
\def\cjaa{Chinese J. Astron. Astrophys.}%
\def\icarus{Icarus}%
\def\jcap{J. Cosmology Astropart. Phys.}%
\def\jrasc{JRASC}%
\def\mnras{MNRAS}%
\def\memras{MmRAS}%
\def\na{New A}%
\def\nar{New A Rev.}%
\def\pasa{PASA}%
\def\pra{Phys.~Rev.~A}%
\def\prb{Phys.~Rev.~B}%
\def\prc{Phys.~Rev.~C}%
\def\prd{Phys.~Rev.~D}%
\def\pre{Phys.~Rev.~E}%
\def\prl{Phys.~Rev.~Lett.}%
\def\pasp{PASP}%
\def\pasj{PASJ}%
\def\qjras{QJRAS}%
\def\rmxaa{Rev. Mexicana Astron. Astrofis.}%
\def\skytel{S\&T}%
\def\solphys{Sol.~Phys.}%
\def\sovast{Soviet~Ast.}%
\def\ssr{Space~Sci.~Rev.}%
\def\zap{ZAp}%
\def\nat{Nature}%
\def\iaucirc{IAU~Circ.}%
\def\aplett{Astrophys.~Lett.}%
\def\apspr{Astrophys.~Space~Phys.~Res.}%
\def\bain{Bull.~Astron.~Inst.~Netherlands}%
\def\fcp{Fund.~Cosmic~Phys.}%
\def\gca{Geochim.~Cosmochim.~Acta}%
\def\grl{Geophys.~Res.~Lett.}%
\def\jcp{J.~Chem.~Phys.}%
\def\jgr{J.~Geophys.~Res.}%
\def\jqsrt{J.~Quant.~Spec.~Radiat.~Transf.}%
\def\memsai{Mem.~Soc.~Astron.~Italiana}%
\def\nphysa{Nucl.~Phys.~A}%
\def\physrep{Phys.~Rep.}%
\def\physscr{Phys.~Scr}%
\def\planss{Planet.~Space~Sci.}%
\def\procspie{Proc.~SPIE}%

\end{document}